\documentclass[pra,aps,twocolumn,showpacs]{revtex4}

\usepackage{graphicx}
\usepackage{bm}

\begin{document}

\title{Gaussification of quantum states of traveling light beams in atomic memory}

\author{Jarom\'{\i}r Fiur\'{a}\v{s}ek}
\affiliation{Department of Optics, Palack\'{y} University, 
17. listopadu 12, 77146 Olomouc, Czech Republic}

\begin{abstract}
We propose and investigate a protocol for Gaussification of quantum states of traveling light beams in an atomic quantum memory that couples to light
via quantum non-demolition interaction. The protocol relies on a periodic switching between two different QND couplings and the total coupling strength scales
only logarithmically with number of Gaussified light modes. The present scheme can be used to prepare entangled states
of two distant atomic ensembles and to purify and Gaussify noisy non-Gaussian entangled states of light while simultaneously storing the purified state in atomic
memories. 
\end{abstract}


\maketitle

\section{Introduction}

Quantum information processing with continuous  variables (CV QIP) exploits the information encoding into 
variables with continuous spectra such as quadrature operators of light modes or collective atomic spin
of mesoscopic atomic ensembles \cite{Cerf07,Braunstein05}. This approach exhibits distinct advantages, such as deterministic preparation
of entangled quantum states of light, highly efficient homodyne detection enabling deterministic quantum teleportation \cite{Furusawa98,Sherson06}
and a suitable interface between light and atoms \cite{Julsgaard01,Julsgaard04,Appel09}. The latter relies on the collective enhancement of the atoms-light 
coupling by means of strong auxiliary 
coherent light beam and a mesoscopic atomic ensemble containing large number of atoms. Under certain circumstances, the atoms-light coupling can be treated 
as a quantum non-demolition (QND) interaction \cite{Kuzmich98,Duan00} that enables deterministic entanglement of two distant atomic ensembles \cite{Julsgaard01} 
and implementation of a quantum memory for light \cite{Julsgaard04}. 

A central role in CV QIP is played by Gaussian quantum states whose Wigner function has Gaussian form. 
On one hand, these states can be easily prepared experimentally using coherent laser beams, passive linear optics and squeezers, 
and on the other hand they admit efficient theoretical description in terms of covariance matrices and displacement vectors. 
Recall that covariance matrix $\gamma$ associated with an $M$-mode state is defined as $\gamma_{jk}=\langle\{ \Delta r_j,\Delta r_k\}\rangle$, where
$\Delta r_j=r_j-\langle r_j\rangle$, $r=(x_1,p_1,\ldots,x_M,p_M)$, and $x_j$ and $p_j$ denote the amplitude and phase quadrature operators  of the $j$th mode, respectively. 
 The quadrature operators  satisfy the canonical commutation relations $[x_j,p_k]=i\delta_{jk}$. Interestingly, the Gaussian states turn 
 out to be extremal in certain sense \cite{Wolf06}. 
Among all states with a given covariance matrix, several important quantities such as distillable secret key rate or certain entanglement measures 
are minimized by a Gaussian state. This observation plays a crucial role in proofs of the security of quantum key distribution schemes with coherent
states and homodyne detection \cite{GarciaPatron06}. An essential ingredient in the proof of extremality of Gaussian states is a specific 
symplectic transformation that produces balanced superpositions of quadratures of all $M$ input modes \cite{Wolf06}.
In particular, for the output mode where all inputs appear with positive weight we have,
\begin{eqnarray}
x&=&\frac{1}{\sqrt{M}}\sum_{j=1}^M x_j, \nonumber  \nonumber \\
p&=&\frac{1}{\sqrt{M}}\sum_{j=1}^M p_j.
\label{Gaussification}
\end{eqnarray}
Here $x$, $p$ represent the amplitude and phase quadrature of the output mode. If all $M$ input modes are uncorrelated and prepared in the same state $\rho$ 
with zero displacement, $\langle x_j\rangle=\langle p_j\rangle=0$, then in the limit $M\rightarrow \infty$ the output state  converges to a Gaussian state 
with the same covariance matrix 
as that of the input states \cite{Wolf06}.  Operation (\ref{Gaussification}) can be implemented by mixing light beams on an array of beam splitters \cite{Paul96,Fiurasek07}. 
The Gaussification procedure can be modified by performing measurements on the other output ports of the beam splitters and conditioning
on the measurement outcomes. This latter approach forms a core part of the entanglement distillation schemes for continuous-variable quantum 
states \cite{Browne03,Eisert04,Fiurasek07b,Hage08,Dong08}.

In this paper we propose and investigate a scheme for Gaussfication of states of traveling light beams mediated by their interaction with atomic quantum memory. 
The procedure is inspired by  protocol for storage of states of light in atomic quantum memory \cite{Julsgaard04} and 
by scheme for coherent-state information concentration in atomic memory \cite{Herec06}. The scheme involves repeated switching between two 
different types of QND coupling which ensures that the total QND coupling strength between light and atoms grows only logarithmically with the number 
of Gaussified light modes $M$. Another appealing feature of the suggested  scheme is that
the light beams need not be perfectly synchronized and can arrive at different times \cite{Herec06} and the resulting Gaussified state
is stored in a memory. This protocol can be used to prepare an atomic memory in an arbitrary Gaussian quantum state. 
It can also be used to purify and distill entangled states of light such that the purified entangled state is transferred from light onto atoms 
and stored in a pair of distant atomic memories.

The rest of the present paper is organized as follows. In Sec. II we will briefly review the quantum atoms-light interface and describe our protocol.
In section III we will investigate how this protocol can be used to  map an entangled Gaussian state of light beams onto two distant atomic memories.
In Sec. IV we will then demonstrate that our scheme can be used for entanglement purification of phase-diffused 
two-mode squeezed states of light. Finally, Sec. V contains conclusions.

\begin{figure}[!t!]
\centerline{\includegraphics[width=0.85\linewidth]{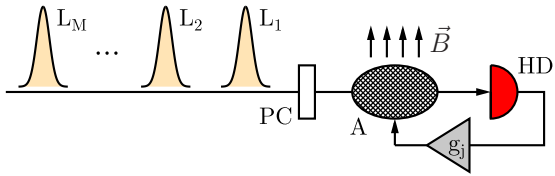}}
\caption{(Color online) Proposed scheme for Gaussification of quantum states of $M$ propagating light modes in atomic quantum memory. Each light beam ($L_j$) 
 interacts with an atomic ensemble (A) and is measured by homodyne detector (HD).
The atomic state is then displaced by an amount proportional to the measurement outcome. 
The QND interaction between light and atoms is periodically  switched between Hamiltonians (\ref{HxLpA}) and (\ref{HpLxA}) with the help of a 
fast polarization controller (PC) and  magnetic field pulses ($\vec{B}$). For more details, see main text.}
\end{figure}

\section{Gaussification in atomic memory}

We consider configuration where light couples to the collective (pseudo) spin of an ensemble of atoms, $\bm{J}$. Its cartesian components satisfy the 
SU(2) commutation relations, 
\begin{equation}
[J_y,J_z]=i J_x.
\label{Jcommutator}
\end{equation}
The atomic spins are initially all aligned along the $x$ axis (e.g. by optical pumping) such that $\langle J_x\rangle= F N_A$,
where $N_A$ is the total number of atoms and $F$ denotes the total ground state angular momentum of a single atom. 
For large $N_A$, $J_x$ can be treated as a classical quantity and the operator replaced by its mean value in the 
commutation relation (\ref{Jcommutator}). We can then define effective atomic quadrature operators $x_A=J_y/\sqrt{\langle J_x\rangle}$ and 
$p_A=J_z/\sqrt{\langle J_x\rangle}$, that approximately satisfy the canonical commutation relations, $[x_A,p_A]=i$. 
Imagine a vertically polarized signal light beam together with an auxiliary strong coherent horizontally polarized light beam propagating through the 
atomic ensemble along the $z$ axis, see Fig.~1. Under certain circumstances, the off-resonant atoms light coupling is governed by a quantum non-demolition 
(QND) interaction \cite{Kuzmich98,Duan00},
\begin{equation}
H=\hbar \kappa x_L p_A,
\label{HQND}
\end{equation}
where $\kappa \propto \sqrt{N_A N_L}$, and $N_L$ is the number of photons in the auxiliary horizontally polarized light beam. Note that alternatively, 
the two polarization modes can be replaced by two spatial modes with  atoms placed in one arm of a Mach-Zehnder interferometer \cite{Appel09}.
In the Heisenberg picture, the QND interaction transforms the atomic and light quadratures as follows,
\begin{eqnarray}
x_{A}^{\mathrm{out}}&=&x_{A}+\kappa x_{L},  \nonumber \\
p_{A}^{\mathrm{out}}&=&p_{A},  \nonumber \\
x_{L}^{\mathrm{out}}&=&x_{L},  \nonumber \\
p_{L}^{\mathrm{out}}&=&p_{L}-\kappa p_{A}. 
\label{QNDtransformation}
\end{eqnarray}
The QND coupling with  $\kappa=1$ can be combined with measurement of the output light  quadrature $p_{L}^{\mathrm{out}}$
and displacement of the atomic quadrature $p_{A}^{\mathrm{out}}$, $p_{A}^{\mathrm{out}} \rightarrow p_{A}^{\mathrm{out}}+gp_{L}^{\mathrm{out}}$, 
 where $g=\frac{1}{2}$ is the displacement gain. In this way we can partly emulate beam splitter operation by QND coupling \cite{Filip08}
and we obtain 
\begin{equation}
x_{A}^{\mathrm{out}}= x_A+x_L, \qquad p_{A}^{\mathrm{out}}= \frac{1}{2}(p_A+p_L).
\end{equation}
A naive $M$-fold repetition of this protocol with $M$ light modes, fixed interaction strength $\kappa=1$ and variable feedback gain $g_k=1/(k+1)$, yields 
\begin{eqnarray}
x_{A}^{\mathrm{out}}&=&x_A+\sum_{j=1}^M x_{L,j},  \nonumber \\
p_{A}^{\mathrm{out}}&=&\frac{1}{M+1}\left(p_A+\sum_{j=1}^M p_{L,j}\right).
\label{naiveoutput}
\end{eqnarray}
This closely resembles the target transformation (\ref{Gaussification}) but the atomic state is additionally squeezed by a factor of $1/\sqrt{M+1}$. This squeezing which 
grows with $M$ would eventually make the state of the memory very fragile and would enhance any imperfections in the feedback displacement applied 
to the squeezed quadrature $p_A$. Another drawback of this approach is that the total squared coupling strength,
\begin{equation}
K_{\mathrm{tot}}^2= \sum_{j=1}^M \kappa_j^2,
\label{Ktot}
\end{equation}
which is proportional to the total number of photons interacting with the atomic ensemble, will increase linearly with $M$, $K_{\mathrm{tot}}^2=M$. 
However, it is desirable to keep $K_{\mathrm{tot}}^2$ as small as possible because certain  decoherence effects in the atomic memory increase with  $K_{\mathrm{tot}}^2$.
The QND coupling  constant $\kappa$ can be expressed as $\kappa^2=d \eta$, where $d$ is resonant optical depth of the atomic ensemble
and $\eta$ is the atomic depumping factor \cite{Hammerer04,Hammerer10}. For a given experimental configuration the optical depth $d$ is fixed and $\kappa$ 
can be increased only by  increasing the number of photons $N_L$ in the auxiliary optical beam, which however simultaneously increases also atomic depumping 
due to scattering of photons.  It was shown in Ref. \cite{Hammerer04} that this decoherence effect can be modeled as a lossy Gaussian channel
with added thermal noise. Covariance matrix of the atomic state after QND coupling further transforms as 
\begin{equation}
\gamma_{A}' = (1-\eta) \gamma_{A}+2\eta  \gamma_{\mathrm{vac}},
\label{decoherence}
\end{equation}
where $\gamma_{\mathrm{vac}}=\mathrm{diag}(1,1)$ is covariance matrix of vacuum noise. Clearly, the attenuation and thermal noise addition described by Eq. (\ref{decoherence}) decreases 
efficiency of atoms-light coupling. To minimize this detrimental effect it is crucial to keep the total squared coupling $K_{\mathrm{tot}}^2$ as small as possible.

If we just reduce the coupling strength while keeping it fixed, $\kappa=1/\sqrt{M}$, then with appropriate gains we obtain \cite{Herec06},
\begin{eqnarray}
x_{A}^{\mathrm{out}}&=&x_A+\frac{1}{\sqrt{M}}\sum_{j=1}^M x_{L,j},  \nonumber \\
p_{A}^{\mathrm{out}}&=& \frac{1}{\sqrt{M}}\sum_{j=1}^M p_{L,j}.
\label{Herecoutput}
\end{eqnarray}
This procedure achieves a fixed total squared coupling strength $K_{\mathrm{tot}}^2=1$ independent of $M$, but the output amplitude quadrature $x_{A}^{\mathrm{out}}$
suffers from an uncompensated noise represented by the term $x_A$ on the right hand side of Eq. (\ref{Herecoutput}). This noise would preclude faithful Gaussification of light states even in the 
asymptotic limit $M \rightarrow \infty$ and would limit the amount of entanglement and purity of the state that can be created in the memory via Gaussification. The noise could be
reduced by a presqueezing of the atomic quantum memory achieved by a QND measurement of the $x_A$ quadrature  with coherent probe light. However, squeezing $x_A\rightarrow x_A/\sqrt{M}$ 
would require  QND coupling $\kappa^2=M-1$, so we recover linear scaling of $K_{\mathrm{tot}}^2$ with $M$.

We now present an alternative  protocol that avoids the unwanted squeezing, residual noise or presqueezing of atomic memory 
and achieves logarithmic scaling of $K_{\mathrm{tot}}^2$ with $M$.
The Gaussification  is still based on repeated interaction 
of the atomic memory with many copies of the light state. However, we adjust the value of the coupling constant $\kappa_M$ in each step
by controlling the intensity of the auxiliary strong coherent laser beam. We also repeatedly switch between two 
different QND couplings, 
\begin{equation}
H_{2N-1}=\hbar \kappa_{2N-1} x_{L,2N-1} p_A,
\label{HxLpA}
\end{equation}
and
\begin{equation}
H_{2N}=-\hbar \kappa_{2N} p_{L,2N} x_A.
\label{HpLxA}
\end{equation}
Switching between the two Hamiltonians (\ref{HxLpA}) and (\ref{HpLxA}) can be performed for instance by rotating the collective atomic spin of the atomic ensemble 
by a magnetic pulse and by rotating the polarization state of the light beam by a fast electrooptical modulator.
The protocol is designed such that after $M$ steps the atomic quadrature operators are given by
\begin{eqnarray}
x_{A,M}&=&\frac{C_{M}}{\sqrt{M+1}}\left(\frac{x_A}{C_0}+\sum_{j=1}^{M} x_{L,j}\right),  \nonumber \\
p_{A,M}&=&\frac{1}{C_{M}\sqrt{M+1}}\left(C_0 p_A+\sum_{j=1}^{M}p_{L,j}\right),
\label{target}
\end{eqnarray}
where $C_0$ is a free parameter and $C_M$ depends on $C_0$ and $M$, as we shall see below. Note that $C_M$ specifies the overall squeezing of the
atomic state after $M$ steps of the Gaussification protocol while $C_0$ represents an additional squeezing of the initial atomic state.

When analyzing the protocol we must distinguish odd and even steps. Consider first an odd step, $j=2N+1$. 
The atomic and light quadratures are transformed according to the interaction Hamiltonian (\ref{HxLpA}),
\begin{eqnarray}
x_{A}^{\mathrm{out}}&=&x_{A,2N}+\kappa_{2N+1}x_{L,2N+1},  \nonumber \\
p_{A}^{\mathrm{out}}&=&p_{A,2N},  \nonumber \\
x_{L}^{\mathrm{out}}&=&x_{L,2N+1},  \nonumber \\
p_{L}^{\mathrm{out}}&=&p_{L,2N+1}-\kappa_{2N+1}p_{A,2N}. 
\end{eqnarray}
The output light quadrature $p_{L}^{\mathrm{out}}$ is measured and the atomic quadrature $p_{A}^{\mathrm{out}}$ is displaced by an amount
of $g_{2N+1}p_{L}^{\mathrm{out}}$, where $g_{2N+1}$ denotes the feedback gain. After $2N+1$ steps the atomic quadratures 
thus read
\begin{eqnarray}
x_{A,2N+1}&=&x_{A,2N}+\kappa_{2N+1}x_{L,2N+1},  \nonumber \\
p_{A,2N+1}&=&(1-g_{2N+1}\kappa_{2N+1})p_{A,2N}+g_{2N+1}p_{L,2N+1}.   \nonumber \\
\end{eqnarray}
In order to preserve the structure of the transformation (\ref{target}) we must set
\begin{eqnarray}
\kappa_{2N+1}&=&\frac{C_{2N}}{\sqrt{2N+1}},   \nonumber \\
 g_{2N+1}&=&\frac{\sqrt{2N+1}}{2(N+1) C_{2N}},
 \label{oddkappa}
\end{eqnarray}
which yields
\begin{equation}
C_{2N+1}=\sqrt{\frac{2N+2}{2N+1}}\,C_{2N}.
\label{Codd}
\end{equation}

\begin{figure}[!b!]
\centerline{\includegraphics[width=0.85\linewidth]{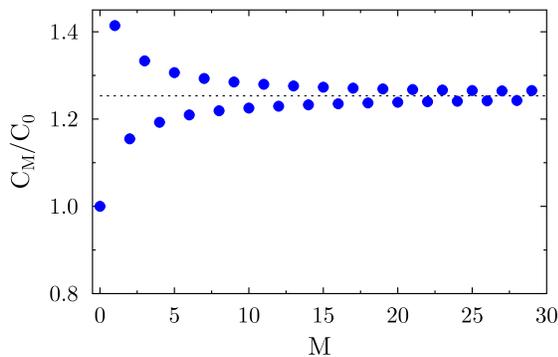}}
\caption{The ratio $C_M/C_0$ is plotted as a function of $M$ (blue filled circles). The horizontal dashed line indicates the asymptotic value
of the ratio which is equal to $\sqrt{\pi/2}\approx 1.253$.}
\end{figure}

We next proceed with an even step of the protocol, $j=2N+2$. In the Heisenberg picture, the coupling (\ref{HpLxA}) gives rise to the transformations,
\begin{eqnarray}
x_{A}^{\mathrm{out}}&=&x_{A,2N+1},  \nonumber \\
p_{A}^{\mathrm{out}}&=&p_{A,2N+1}+\kappa_{2N+2}p_{L,2N+2},  \nonumber \\
x_{L}^{\mathrm{out}}&=&x_{L,2N+2}-\kappa_{2N+2}x_{A,2N+1},  \nonumber \\
p_{L}^{\mathrm{out}}&=&p_{L,2N+2}. 
\end{eqnarray}
We measure the output light quadrature $x_{L}^{\mathrm{out}}$ and displace the atomic quadrature $x_{A}^{\mathrm{out}}$ by
$g_{2N+2}x_{L}^{\mathrm{out}}$. If we choose
\begin{eqnarray}
\kappa_{2N+2}&=&\frac{1}{C_{2N+1} \sqrt{2N+2} },   \nonumber \\
 g_{2N+2}&=&\frac{\sqrt{2N+2}}{2N+3}\, C_{2N+1},
 \label{evenkappa}
\end{eqnarray}
then we preserve the structure of the target transformation (\ref{target}) with
\begin{equation}
C_{2N+2}=\sqrt{\frac{2N+2}{2N+3}}\,C_{2N+1}.
\label{Ceven}
\end{equation}
By combining Eqs. (\ref{Codd}) and (\ref{Ceven}) we obtain a recurrence formula for $C_{2N}$,
\begin{equation}
C_{2N+2}=\frac{2N+2}{\sqrt{(2N+3)(2N+1)}}C_{2N},
\end{equation}
which yields,
\begin{equation}
C_{2N}=\sqrt{2N+1}\,\frac{(2^N N!)^2}{(2N+1)!} \, C_0.
\end{equation}

The ratio $C_M/C_0$ is plotted in Fig.~2. We can see that this ratio lies in a narrow interval, $C_0 \leq C_M \leq \sqrt{2} C_0$ and it quickly approaches
a fixed asymptotic value, 
\begin{equation}
\lim_{M\rightarrow \infty} \frac{C_M}{C_0}=\sqrt{\frac{\pi}{2}}.
\end{equation}
This indicates that the squeezing of the atomic state remains bounded and is saturated at a value of $C_\infty=\sqrt{\pi/2}\, C_0$. In particular,
if we set $C_0=\sqrt{2/\pi}$ then the transformation (\ref{target}) becomes in the limit of infinite $M$ exactly equivalent to the target operation (\ref{Gaussification}).
For any finite $M$ we can choose $C_0$ such that $C_M=1$ will hold. In this case, the initial atomic quadratures will appear in the superposition
rescaled by the squeezing factor $C_0$. The optimal choice of $C_0$ will generally depend on the intended application of the protocol and on $M$.

\begin{figure}[!t!]
\centerline{\includegraphics[width=0.85\linewidth]{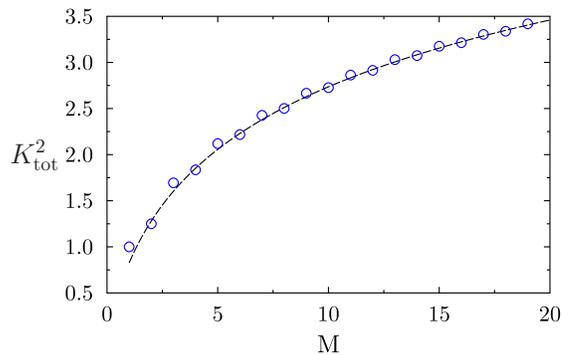}}
\caption{Dependence of the total coupling strength $K_{\mathrm{tot}}^2$ on the number of steps $M$ of the protocol is plotted for $C_0=1$ (blue empty circles). 
The dashed curve represents the best fit by a logarithmic function.}
\end{figure}

The asymptotically constant value of $C_M/C_0$ implies that the QND coupling strength scales as $\kappa_M \propto C_0/\sqrt{M}$. The total squared 
coupling strength $K_{\mathrm{tot}}^2$ defined by Eq. (\ref{Ktot}) thus increases only logarithmically with $M$. This is graphically illustrated in Fig.~3 where the dependence
of $K_{\mathrm{tot}}^2$ on $M$ is plotted for $C_0=1$ together with the best logarithmic fit $K_{\mathrm{tot}}^2\approx 0.05+1.12\ln(M+1)$.

\section{Mapping of Gaussian states of light onto atoms}

If the Gaussification protocol is applied to light beams prepared in identical independent Gaussian states with covariance matrix $\gamma_L$ 
and zero displacement, then the protocol maps the Gaussian state of light into atomic memory. Assuming that the atomic memory is initially prepared
in some Gaussian state with covariance matrix $\gamma_A$ and zero displacement, the covariance matrix of a Gaussian state in atomic quantum memory
after $M$ steps of the protocol reads
\begin{equation}
\gamma_{A}^{(M)}=\frac{1}{M+1}S_A\gamma_A S_A^T+\frac{M}{M+1} S_L\gamma_L S_L^T,
\end{equation}
where 
\begin{equation}
S_A=\left(
\begin{array}{cc}
 \frac{C_M}{C_0} & 0  \nonumber \\
 0 & \frac{C_0}{C_M}
\end{array}
\right),
\qquad
S_L=\left(
\begin{array}{cc}
 C_M & 0  \nonumber \\
 0 & C_M^{-1}
\end{array}
\right),
\end{equation}
 represent the effective squeezing of the atomic and light contributions.
In the limit of large $M$ and by choosing $C_0=\sqrt{2/\pi}$ such that $\lim_{M\rightarrow \infty} C_M=1$ the many copies of a Gaussian state of light are mapped onto a single copy stored in
the memory, 
\begin{equation}
\lim_{M\rightarrow \infty}\gamma_A^{(M)}=\gamma_L.
\end{equation}

\begin{figure}[!b!]
\centerline{\includegraphics[width=0.9\linewidth]{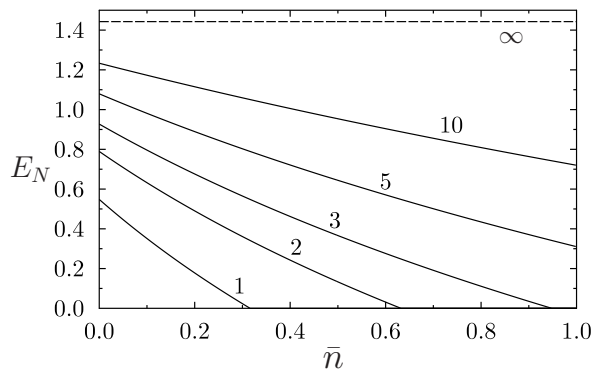}}
\caption{Entanglement of two atomic ensembles created by local mappings of parts of pure two-mode squeezed vacuum onto the atomic memories. The logarithmic negativity
$E_N$ of the two-mode atomic state is plotted as a function of the initial mean number of thermal quanta $\bar{n}$ in the memory. This dependence is 
shown  for several number of Gaussification steps $M=\{1,2,3,5,10\}$ specified by numerical labels. The dashed line indicates 
the maximum entanglement achievable in the asymptotic limit $M\rightarrow \infty$, squeezing constant $r=0.5$ and $C_0=1$.} 
\end{figure}

This procedure can be used to establish entanglement between two distant atomic quantum memories by mapping parts of an entangled state of light 
into them. The Gaussification procedure is applied locally to each atomic memory and sequence of light modes. 
We shall assume that  both atomic quantum memories are initially in a thermal state with mean number of quanta $\bar{n}$ and that the 
light modes are prepared in a pure two-mode squeezed vacuum state $|\Psi_{\mathrm{TMS}}\rangle$ with squeezing constant $r$ and covariance matrix
\begin{equation}
\gamma_L=\left(
\begin{array}{cccc}
\cosh(2r) & 0 & \sinh(2r) & 0  \nonumber \\
0 & \cosh(2r) & 0 & - \sinh(2r)  \nonumber \\
\sinh(2r) & 0 & \cosh(2r) & 0  \nonumber \\
0 & -\sinh(2r) & 0 & \cosh(2r)  
\end{array}
\right).
\end{equation}
The entanglement between two quantum memories can be conveniently quantified by a logarithmic negativity $E_N$, which is an analytically computable measure
of entanglement for two-mode Gaussian states \cite{Vidal02}. We have
\begin{equation}
E_{N}=\max(0,-\log_2(\mu)),
\end{equation}
where $\mu$ is the lower symplectic eigenvalue of a two-mode covariance matrix corresponding to a partially transposed state of the two atomic memories,
\begin{equation}
\mu^2=\frac{M^2e^{-4r}+M(2\bar{n}+1)\left(C_0^2+C_0^{-2}\right)e^{-2r}+(2\bar{n}+1)^2}{(M+1)^2} .
\end{equation}
The entanglement created in the atomic memory is maximized when the symplectic eigenvalue $\mu$ is minimized. This occurs for $C_0=1$
and we have
\begin{equation}
\mu_{\mathrm{opt}}= \frac{Me^{-2r}+2\bar{n}+1}{M+1}.
\end{equation}
The dependence of $E_N$ on $\bar{n}$ is plotted in Fig. $4$ for various numbers of steps $M$ of the protocol.
We can see that each step of the protocol increases entanglement of the two memories. For a fixed number of steps the amount of created
entanglement monotonically decreases with increasing amount of thermal noise $\bar{n}$.

\begin{figure}[!t!]
\centerline{\includegraphics[width=0.9\linewidth]{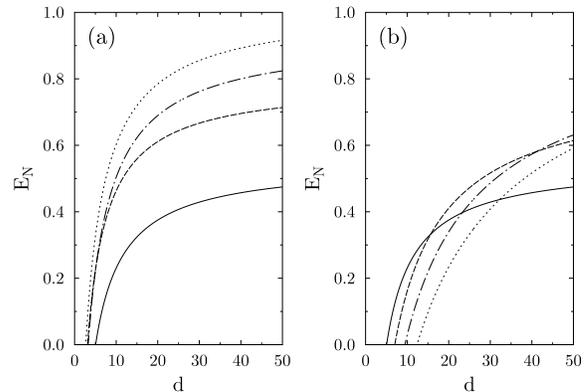}}
\caption{The same as Fig. 4 but the logarithmic negativity
$E_N$ of the two-mode atomic state is plotted as a function of the resonant optical depth $d$ of the atomic ensemble. The various lines correspond to different 
number of Gaussification steps $M=1$ (solid line), $M=2$ (dashed line), $M=3$ (dashed-dotted line) and $M=4$ (dotted line). 
Shown are results for protocol with variable coupling strength $\kappa$ (a), and for a protocol with fixed $\kappa=1$ (b).
The parameters read $r=0.5$, $\bar{n}=0$, and $C_0=1$.
} 
\end{figure}

So far we have considered an idealized scenario with no decoherence of the atomic memory.
We now investigate the decoherence effects caused by finite resonant optical depth $d$ and nonzero atomic depumping $\eta$.
In particular, we compare the Gaussification protocol suggested in this paper with the simple scheme with fixed coupling strength as described by Eq. (\ref{naiveoutput}).
These two protocols are both used to prepare two-mode squeezed vacuum state of two atomic ensembles by mapping many copies
of a two-mode squeezed vacuum state of light onto atoms. In a limit of infinite $d$ and zero $\eta$, both protocols are completely equivalent and create 
the same amount of entanglement after $M$ Gaussification steps. However, the atomic decoherence influences the two schemes in a different way, because they 
exhibit different $K_{\mathrm{tot}}^2$. We have numerically calculated covariance matrix of a two-mode atomic state after M steps of each protocol. In the numerical 
simulation, the decoherence map (\ref{decoherence}) with appropriate depumping factor $\eta_j=\kappa_j^2/d$ is applied to a covariance matrix of the atomic state 
after each interaction of atoms with light.  The results  are reported in Fig. 5 where we plot the logarithmic negativity of the atomic state as a function
of $d$ for several steps of the protocol, $M=1,2,3,4$. A comparison of Figs. 5(a) and 5(b) reveals that
the protocol exploring switching between two different QND couplings and variable coupling strength can generate substantially more entanglement  
than the protocol with fixed $\kappa$. In this latter case already the second step can be in fact detrimental and lead to reduction of atomic 
entanglement due to decoherence if $d$ is small enough.

\begin{figure}[!t!]
\centerline{\includegraphics[width=0.98\linewidth]{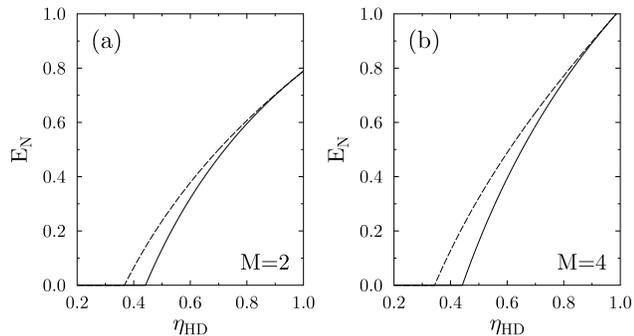}}
\caption{The same as Fig. 4 but logarithmic negativity of the atomic state $E_N$ after $2$ (a) and $4$ (b) steps of the protocol is plotted as a function 
of homodyne detection efficiency $\eta_{\mathrm{HD}}$. The results are shown for 
protocol with variable coupling strength (solid line) and for a protocol with fixed $\kappa=1$ (dashed line). The parameters read $r=0.5$,  $\bar{n}=0$, and 
$C_0=1$.} 
\end{figure}

Besides atomic depumping, another effect that can negatively influence the Gaussification procedure is imperfect homodyne detection that is used to measure 
the output light quadrature.  
The quadrature operator detected by an imperfect homodyne detector with detection efficiency $\eta_{\mathrm{HD}}$ reads 
$p_{L,\mathrm{eff}}=\sqrt{\eta_{\mathrm{HD}}}p_{L}+\sqrt{1-\eta_{\mathrm{HD}}}p_{L,\mathrm{noise}}$ where $p_{L,\mathrm{noise}}$
represents quadrature of an auxiliary vacuum state. In the feedback, we can compensate for $\eta_{\mathrm{HD}}<1$ by re-scaling the gain, $g \rightarrow g/\sqrt{\eta_{\mathrm{HD}}}$.
The noise added to an atomic quadrature by feedback is then given by $g\sqrt{\frac{1-\eta_{\mathrm{HD}}}{\eta_{\mathrm{HD}}}}p_{L,\mathrm{noise}}$.
This noise increases both with decreasing efficiency of homodyne detection and with  increasing feedback gain. 
The  protocol with fixed $\kappa$ specified by Eq. (\ref{naiveoutput}) is less sensitive to this decoherence mechanism, because there the gain 
scales with number of steps $M$ as $1/M$ while for the scheme with varying $\kappa$ the scaling is  $1/M^{1/2}$. This is confirmed by numerical calculations whose results
are shown in  Fig. 6. In order to unambiguously assess the influence of $\eta_{\mathrm{HD}}$, we have neglected other decoherence mechanisms and considered the limit of $d\rightarrow \infty$.
We can see that the entanglement generated in the atomic memory decreases with decreasing $\eta_{\mathrm{HD}}$. However, both  protocols are  rather robust and atomic entanglement 
can be generated even if $\eta_{\mathrm{HD}}<50\%$ while typical efficiency of homodyne detection can exceed $90\%$.
Note also that in the considered example the difference between the two protocols becomes non-negligible only for homodyne detection efficiency less than 80\%.

\section{Entanglement Purification}

As a final application of the Gaussification protocol we consider purification and distillation of non-Gaussian entanglement of light beams coupled to atomic memories
 placed at two distant locations A and B.
The protocol is again based on local interactions of parts of entangled states with atomic memory. Additionally, we also make use of 
conditioning on the outcomes of homodyne measurements on light beams after they interacted with the atomic memory \cite{Fiurasek07,Hage08}.
The homodyne detection thus plays a dual role here: on one hand it provides the information required for a feedback onto atoms and on the other
hand it heralds the success of purification. In particular, the scheme succeeds if all measurement outcomes $Q_j$ are sufficiently close to zero, 
$|Q_j|\leq Q_T$, where $Q_T$ is some threshold. In what follows we will consider limit of a very narrow acceptance window which can be mathematically modeled
as conditioning on $Q_j=0$. Note that the acceptance condition must be simultaneously satisfied at both locations A and B. The protocol therefore requires classical communication 
between A and B which is used to establish success or failure. Although the present scheme involves quantum memories, it is not a quantum repeater
because the memories do not improve the scaling of the success probability of the protocol. 
However, the quantum memories do enable purification and Gaussification of many copies of entangled states of light emitted at different times \cite{Herec06}.

As an explicit example we consider purification of phase-diffused two-mode squeezed states which has previously been successfully demonstrated experimentally 
for  traveling light beams \cite{Hage08}. Phase diffusion arises when the entanglement is distributed over
quantum channels with fluctuating optical length. We assume that the source is located at a center between A and B such that
both modes suffer from random phase shifts $\phi_A$ and $\phi_B$. The phase noise converts the initial pure entangled Gaussian two-mode squeezed 
state $|\Psi_{\mathrm{TMS}}\rangle_{AB}$ into a mixed non-Gaussian state
\begin{equation}
\rho_{AB}=\int \int ( U_A\otimes U_B ) \Psi_{AB} ( U_A^\dagger \otimes U_B^\dagger ) P(\phi_A)  P(\phi_B) d \phi_A d\phi_B,
\end{equation}
where $\Psi_{AB}=|\Psi_{TMS}\rangle\langle \Psi_{TMS}|$, $U_j=\exp(-i \phi_j n_j)$, $n_j$ denotes photon number operator of mode $j$, 
and $P(\phi_j)$ is the probability distribution of random phase shift. In what follows we shall assume Gaussian distribution of random phase shift \cite{Fiurasek07,Hage08},
\begin{equation}
P(\phi_j)=\frac{1}{\sqrt{2\pi\sigma^2}} \exp\left(-\frac{\phi^2}{2\sigma^2}\right),
\end{equation}
where $\sigma$ quantifies the strength of fluctuations. Note that Gaussian distribution of $\phi$ gives rise to highly non-Gaussian $\rho_{AB}$.
For strong enough phase noise, the Gaussian entanglement is completely lost and the two-mode state does not exhibit any quadrature squeezing. In such case
the deterministic Gaussification protocol discussed in previous section is useless, because it would produce a separable state of the two memories at stations A and B. 
In contrast, the protocol augmented by conditioning on outcomes of homodyne detection is capable of converting the initially non-Gaussian entanglement into Gaussian one. 
We quantify the performance of the protocol by logarithmic negativity $E_N$ of the distilled state $\rho$, purity of the distilled state 
$\mathcal{P}=\mathrm{Tr}(\rho^2)$, and total variance 
\begin{equation}
\mathcal{I}=\frac{1}{2}\left(\langle(\Delta x_A-\Delta x_B)^2\rangle+ \langle(\Delta p_A+\Delta p_B)^2\rangle\right).
\end{equation}
The state is entangled if $\mathcal{I}<1$. We also calculate Gaussianity of the state, $G$, which is defined as 
fidelity of the state with a Gaussian state with the same covariance matrix and displacement. It holds that $G \leq 1$ and $G=1$ 
only if the state is Gaussian.

\begin{figure}[!b!]
\centerline{\includegraphics[width=0.98\linewidth]{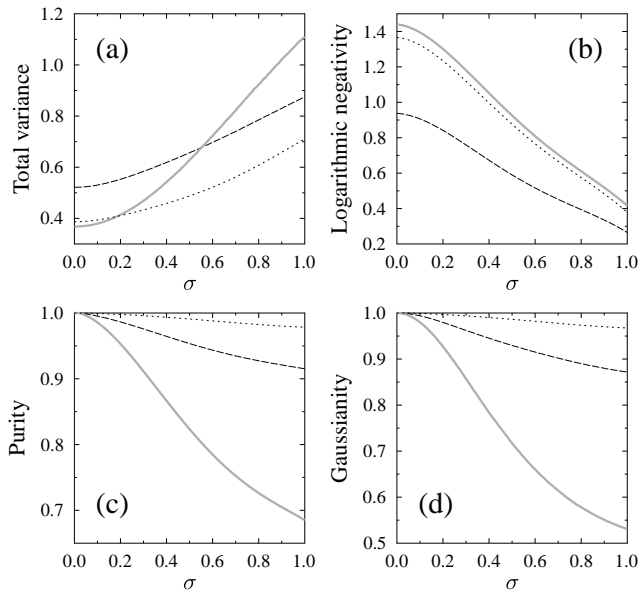}}
\caption{Gaussification of phase-diffused two-mode squeezed vacuum into two distant atomic memories. The figure shows total variance (a), 
logarithmic negativity (b), purity (c), and Gaussianity (d) of the atomic state after 2 (dashed line) and 20 (dotted line) steps of the protocol
 as functions of phase fluctuations strength $\sigma$. For comparison, 
 solid gray line shows the values corresponding to the initial phase-diffused two-mode squeezed state of light. The atomic ensembles are initially
in vacuum state, the two-mode squeezing constant $r=0.5$ and $C_0=\sqrt{2/\pi}$.} 
\end{figure}

The results of numerical simulations are shown in Fig. 7, which shows the total variance, logarithmic negativity, purity and Gaussianity of the state in atomic
memories after two and twenty steps of the protocol. For comparison, the thick solid gray lines indicate values of the considered quantities attained by the 
de-phased light state. Each successful step of the protocol increases purity and Gaussianity and after $20$ steps of the protocol the state 
is very pure and very close to Gaussian state for all $\sigma \leq 1$. Each successful measurement step also increases logarithmic negativity of the 
atomic state and decreases its total variance. For large $\sigma$ the de-phased state of light exhibits $\mathcal{I}>1$, i.e. no Gaussian entanglement.
The purification reduces the total variance such that $\mathcal{I}<1$, hence it converts non-Gaussian entanglement into Gaussian one. 
In the present example, the entanglement stored in atomic memories does not exceed the initial entanglement 
of the de-phased light beam c.f. Fig. 7(b). One reason for this saturation effect is that the two atomic ensembles are initially in a separable
state, in contrast to entanglement distillation protocols with light beams where all copies are initially in the same entangled state \cite{Browne03,Eisert04,Fiurasek07}.
Secondly, the procedure somewhat differs from the previously studied Gaussification-based entanglement distillation schemes 
due to switching between measurements of amplitude and phase quadratures.

\begin{figure}[!t!]
\centerline{\includegraphics[width=0.98\linewidth]{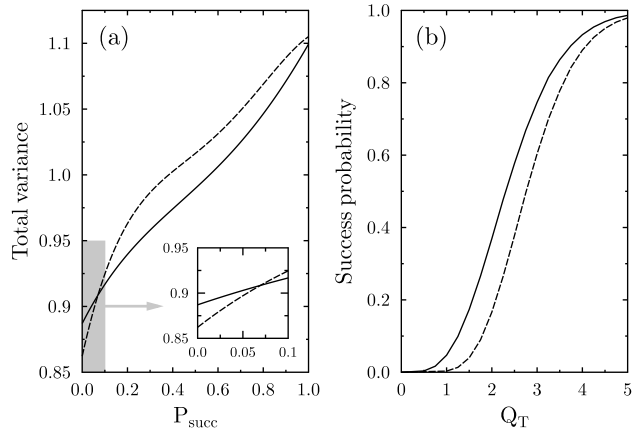}}
\caption{Entanglement purification of phase-diffused two-mode squeezed states with non-zero acceptance threshold $Q_T>0$. 
Figure (a) shows the total variance of atomic state as a function of the overall success probability of the protocol. In figure (b)  we plot
the dependence of overall success probability on acceptance threshold $Q_T$. 
The results are shown for a two-step protocol (solid line) and for a four-step protocol (dashed line). The parameters read $\sigma=1$, $r=0.5$ and $C_0=\sqrt{2/\pi}$.}
\end{figure}

An important parameter of the entanglement purification protocol is its success probability $P_{\mathrm{succ}}$.  In order to assess how the performance of the 
scheme scales with $P_{\mathrm{succ}}$ we have performed numerical simulations of a full protocol with non-zero acceptance threshold $Q_T$ and determined
the dependence of total variance $\mathcal{I}$ of the state of two atomic ensembles on the overall success probability of the protocol. The calculations were performed
for two-step and four-step scheme.  Results reported in Fig. 8 clearly demonstrate that the probabilistic entanglement purification protocol can significantly 
outperform the deterministic scheme for an overall success probability of $10\%$ or even higher. This is in full agreement with previous theoretical 
and experimental results on distillation of phase diffused squeezed and entangled states of light where also good performance for large success 
rates was observed \cite{Hage08,Hage10}.

\section{Conclusions}

In summary, we have proposed and analyzed a protocol for Gaussification of states of $M$ traveling light beams in an atomic memory. 
The protocol is designed for a memory that interacts with light via quantum non-demolition coupling. The scheme involves repeated switching between
$x_Ap_L$ and $p_A x_L$ couplings, homodyne detection of output light and feedback on atoms. In contrast to other proposals, the total coupling strength 
required by this procedure scales only logarithmically with the number of Gaussified modes and no presqueezing operation is required on atoms or light.
All  components of the proposed protocol have already been successfully experimentally demonstrated 
in the past \cite{Julsgaard04,Sherson06} so small-scale demonstrations with $M=2$ or $M=3$ should be feasible with present-day technology.
If combined with conditioning on the outcomes of homodyne detection on output light, the present protocol can be used to purify and Gaussify non-Gaussian mixed entangled states.
This procedure can thus for instance increase performance of certain entanglement-based continuous variable quantum key distribution schemes that require high purity 
and Gaussianity of shared entangled state for optimal performance.

\acknowledgments
The author would like to thank R. Filip for stimulating discussions.
This work was supported by MSMT under projects LC06007, MSM6198959213, and 7E08028, and also by
the EU under the FET-Open project COMPAS (212008).

\end{document}